\documentclass[%
 aip,
 amsmath,amssymb,
 reprint,%
]{revtex4-1}

\usepackage{graphicx}% Include figure files
\usepackage{dcolumn}% Align table columns on decimal point
\usepackage{bm}% bold math
\usepackage[utf8]{inputenc}
\usepackage[T1]{fontenc}
\usepackage{mathptmx}
\usepackage{etoolbox}
\usepackage{xcolor}

\makeatletter
\def\@email#1#2{%
 \endgroup
 \patchcmd{\titleblock@produce}
  {\frontmatter@RRAPformat}
  {\frontmatter@RRAPformat{\produce@RRAP{*#1\href{mailto:#2}{#2}}}\frontmatter@RRAPformat}
  {}{}
}%
\makeatother
\begin{document}

\preprint{AIP/123-QED}

\title[]{Strong effect of fluid rheology on electrokinetic instability and subsequent mixing phenomena in a microfluidic T-junction}
% Force line breaks with \\
\author{F. Hamid}
\author{C. Sasmal*}%
 \email{csasmal@iitrpr.ac.in}
\affiliation{ 
Soft Matter Engineering and Microfluidics Lab, Department of Chemical Engineering, Indian Institute of Technology Ropar, Punjab, India-140001.
}

\date{\today}% It is always \today, today,
             %  but any date may be explicitly specified

\begin{abstract}
When two fluids of different electrical conductivities are transported under the influence of an electric field, the electrokinetic instability (EKI) phenomenon often occurs in a microfluidic device once the electric field strength and conductivity gradient exceed a critical value. This study presents a detailed investigation of how the rheological behaviour of fluid could influence this EKI phenomenon in a microfluidic T-junction. The non-Newtonian power-law model with different values of the power-law index $(n)$ is used to obtain fluids of different rheological behaviours. We find that as the fluid rheological behaviour changes from shear-thickening $(n > 1)$ to shear-thinning $(n < 1)$ via the Newtonian $(n = 1)$ one, the EKI phenomenon is significantly influenced under the same conditions. In particular, the intensity of this EKI phenomenon is found to be significantly higher in shear-thinning fluids than in Newtonian and shear-thickening fluids. As a result, the corresponding mixing phenomenon, often achieved using this EKI phenomenon, is also notably enhanced in shear-thinning fluids compared to that achieved in Newtonian and shear-thickening fluids. A detailed analysis of both the flow dynamics and mixing phenomena in terms of streamlines, velocity fluctuations, concentration field, mixing efficiency, etc., is presented and discussed in this study. We also employ the data-driven dynamic mode decomposition (DMD) technique to analyze the flow field in more detail. In particular, the information on the coherent flow structures obtained with different values of the power-law index facilitates the understanding of both the EKI-induced chaotic convection and mixing phenomena in a better way; for instance, why the mixing efficiency is higher in shear-thinning fluids than that in Newtonian and shear-thickening fluids. Moreover, we observe that the spatial expanse and intensity of these coherent structures differ significantly as the power-law index changes, thereby providing valuable insights into certain aspects of the underlying flow dynamics that otherwise are not clearly apparent from other analyses.
\end{abstract}

\maketitle

\section{\label{sec:level1}Introduction}
Many fluids, such as emulsions, suspensions, polymer solutions, etc., are frequently encountered in micro and nanofluidic systems for further processing and applications~\cite{Anna2008,mei2022editorial,nghe2011microfluidics}. Furthermore, many biofluids, such as blood, saliva, DNA and protein suspensions, cerebrospinal fluid, suspensions of cells and bioparticles, etc., are also often processed in many micro total analysis systems $(\mu$TAS) for chemical and biochemical analyses and detection. All these fluids exhibit various complex non-Newtonian behaviours, for example, shear-thinning, shear-thickening, viscoplasticity, viscoelasticity, etc., instead of showing a simple Newtonian one~\cite{beris2021recent,haward2011extensional,juarez2011extensional,bloomfield1998effects}. Due to the existence of nonlinearity in their rheological behaviours, the underlying physics associated with their transport processes is much more complex than that seen in linear Newtonian fluids. In various micro total analysis systems $(\mu$TAS), these complex fluids are often preferred to transport with the help of the electrokinetic (EK) mechanism-based microdevices than the traditional pressure-driven microdevices, for instance, syringe pumps. This is mainly due to the following reasons: i) the EK-based microdevices do not have any moving parts as they rely on applying an electric field, and hence, they are relatively easy to handle in small-scale microsystems. ii) the EK-based microdevices generate nearly a plug-like velocity profile, thereby offering less flow resistance~\cite{masliyah1994electrokinetic}. As a result, over the years, extensive research efforts have been spent on the development of extensive EK-based microdevices for various purposes, such as transportation and mixing~\cite{li2004electrokinetics}.             

All these EK-based microdevices are based on the fundamental principles of the electrohydrodynamics phenomenon. One such phenomenon is electroosmotic, wherein fluid flow happens due to the formation of an electric double layer (EDL) along the charged surface of a system under the influence of an electric field~\cite{masliyah1994electrokinetic}. Many studies have been conducted in the literature on this electroosmotic phenomenon by considering both simple Newtonian and complex non-Newtonian behaviours of a fluid. For instance, Zhao et al.~\cite{zhao2008analysis} obtained analytical solutions to investigate this phenomenon in power-law fluids flowing in a slit microchannel. They observed a more plug-like velocity profile and an enhancement in the flow rate for pseudoplastic fluids (with power-law index $n \le 1$) than that for Newtonian fluids under the same conditions. A similar investigation for a cylindrical microchannel was carried out in another study by Zhao and Chun~\cite{zhao2013electroosmotic}. Their analysis revealed that the Helmholtz-Smoluchowski velocity of power-law fluids in cylindrical microchannels became dependent on the channel radius, whereas it was independent for a planar surface. Vasu and De~\cite{vasu2010electroosmotic} conducted an investigation for a rectangular microchannel at high zeta potentials and again observed an increase in the volumetric flow rate for pseudoplastic fluids as compared to Newtonian and dilatant fluids (with power-law index $n \ge 1$) as observed by Zhao et al.~\cite{zhao2008analysis} for a slit microchannel. The corresponding study for an elliptical microchannel was carried out by Srinivas~\cite{srinivas2016electroosmotic}. A reduction in the volumetric flow rate was found in an elliptical microchannel compared to a circular one, which was again more significant for shear-thickening fluids compared to shear-thinning fluids. In a recent study, Mehta et al.~\cite{mehta2022enhanced} conducted extensive numerical simulations of electrokinetic mixing of power-law fluids in a non-uniformly charged micromixer with obstacles either arranged in a staggered or an aligned manner. The mixing efficiency in this geometry was greatly influenced by the rheological behaviour of a fluid depending upon the values of the Debye parameter and wall zeta potential. 

Therefore, it can be seen that the rheological behaviour of fluid could significantly influence the electroosmotic phenomenon in microdevices~\cite{zhao2013electrokinetics}. Studies also found that this rheological behaviour of fluid could also regulate the transport of charged particles (electrophoresis phenomenon) to a great extent, which is another electrohydrodynamics phenomenon encountered in many EK-based microdevices~\cite{lee2004electrophoresis,khair2012coupling,hsu2006electrophoresis,hsu2007effect,yeh2009electrophoresis}. Apart from these two electrokinetic phenomena, electrokinetic instability (EKI) is another well-known phenomenon originated due to the presence of an electrical conductivity gradient in the samples during their electrokinetic transport. The gradient in the electrical conductivity in the samples can sometimes occur intentionally, such as in the sample stacking processes, or unintentionally during the handling of multi-dimensional assays. Therefore, like the electroosmotic and electrophoresis phenomena, a reasonable number of studies have also been conducted to understand this electrohydrodynamics instability phenomenon~\cite{lin2009electrokinetic}. For instance, Lin et al.~\cite{lin2004instability} conducted extensive theoretical, numerical, and experimental investigations on this EKI phenomenon in a rectangular microfluidic channel. An unstable and chaotic flow field was seen inside the channel as the applied electric field exceeded a critical value. This results in rapid mixing inside the device. Many other studies also observed such enhancement in the mixing phenomena due to this instability phenomenon in other geometries like Y-shaped microchannel~\cite{jin2010mixing}, T-shaped microdevice with an outlet microchannel having different grooved shapes~\cite{park2005application}, cross-shaped microchannel~\cite{huang2006application}, etc. Chen et al.~\cite{chen2005convective} extensively investigated this phenomenon in a microfluidic T-junction device. They also found unstable and propagating upstream waves in the flow field. Furthermore, they carried out a linear stability analysis to analyze these unstable modes in more detail. The corresponding study in a cross-shaped microchannel with three merging inlets and one outlet was carried out by Posner et al.~\cite{posner2006convective}.

Although a reasonable number of studies are present on the EKI phenomenon in different microdevices; however, most of those studies have dealt with simple Newtonian fluids. There is hardly any study present on how the rheological behaviour of fluid can influence this phenomenon, although a significant number of studies are available on this aspect for other electrohydrodynamics phenomena such as electroosmotic and electrophoresis, as reviewed in some of those studies above. Among very few studies,  Song et al.~\cite{song2019electrokinetic} investigated the influence of fluid viscoelasticity on these EKI instabilities in a T junction microchannel. They found that the fluid viscoelasticity dramatically changes the critical electric field value at which these instabilities emerge in the system. Furthermore, they observe a significant reduction in the speed and frequency of the convective waves originated due to these instabilities in viscoelastic fluids compared to Newtonian fluids. In a very recent numerical study, Sasmal~\cite{Sasmal} obtained the same observations and provided a detailed explanation for this behaviour in viscoelastic fluids, which was missing in the experimental analysis of Song et al.~\cite{song2019electrokinetic}. Apart from these two studies carried out for viscoelastic fluids, there is no other study available on this EKI phenomenon in non-Newtonian fluids, particularly in generalized Newtonian fluids (GNF), such as power-law fluids, even though shear-thinning and shear-thickening behaviours are the most common non-linear behaviours seen in complex fluids. Therefore, one of the major aims of this present study is to investigate how these behaviours of complex fluids can influence the EKI phenomenon and the subsequent mixing phenomenon in a T microfluidic junction.

Furthermore, the present study plans to use the dynamic mode decomposition (DMD) technique to understand how the fluid rheological behaviour is going to influence the underlying coherent flow structures in this geometry due to the EKI phenomenon. This technique is one of the reduced order modeling (ROM) techniques widely used in distilling  the important spatial features of a flow field in terms of
the so-called ‘modes'~\cite{kutz2016dynamic,wu2021challenges,schmid2022dynamic}. This method has already proved its potential to analyze the coherent flow structures in different fluid dynamical problems~\cite{schmid2011applications,schmid2010dynamic}. In fact, it has also been used for the analysis of the EKI phenomenon in simple Newtonian fluids~\cite{dubey2017coherent}. Therefore, another goal of this study is to use the DMD technique to understand in a better way the differences in the coherent flow structures and the mixing phenomenon arising from the EKI phenomenon in Newtonian and non-Newtonian power-law fluids.

\section{\label{sec:GovEqs}Problem statement and governing equations}
\begin{figure}
    \centering
    \includegraphics[trim=7.2cm 5cm 15.5cm 2.5cm,clip,width=10cm]{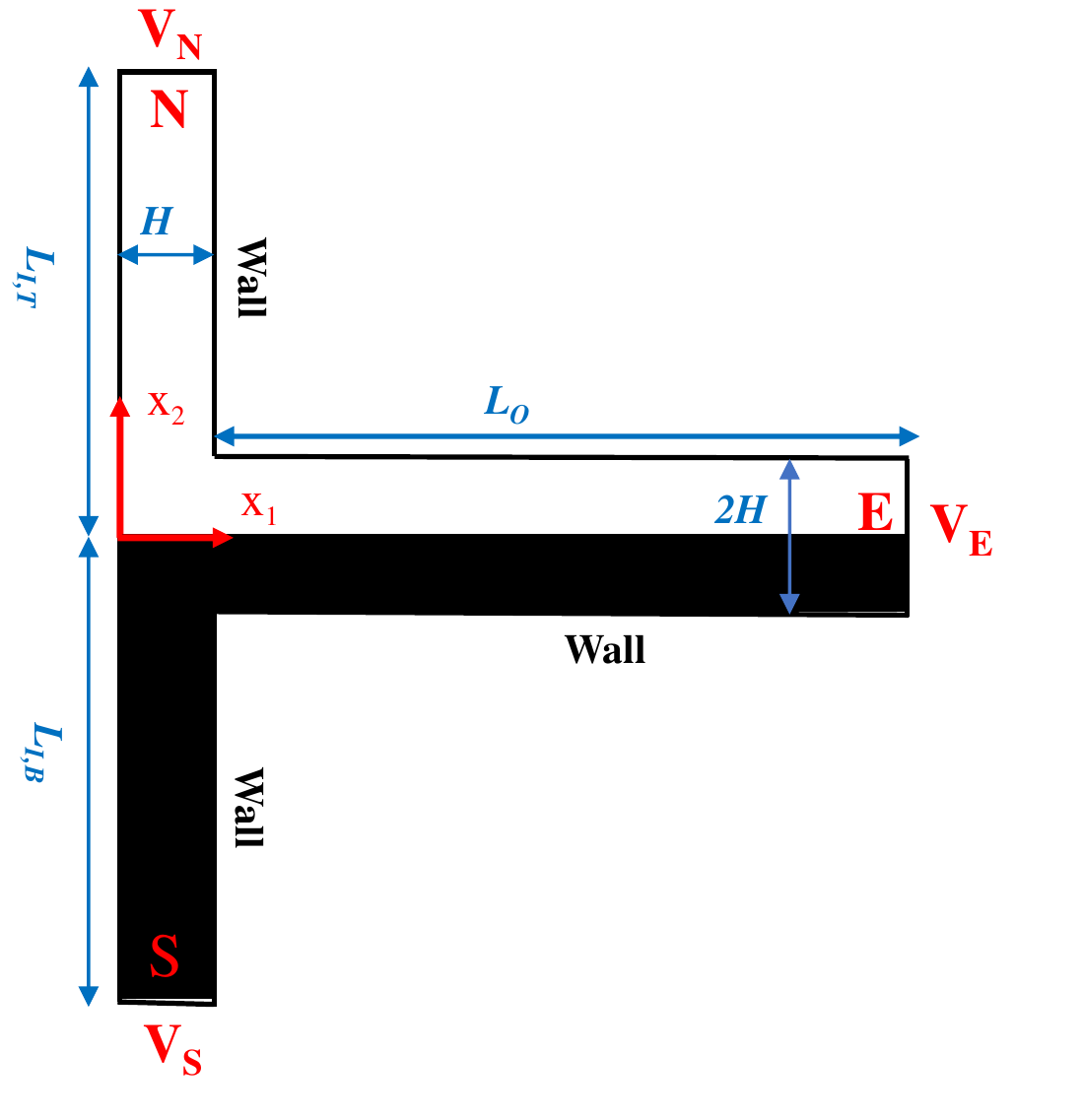}
    \caption{Schematic of the two-dimensional microfluidic T-junction used in this study. Here $2H$ and $H$ are the heights of the outlet and inlet sections of the device, $L_{I,T} = 6.25H$ and $L_{I,B} = 6.25H$ are the lengths of the converging inlets and $L_{O} = 21H$ is the outlet length of the device. $V_{N}$, $V_{S}$, and $V_{E}$ are the applied voltages at the north $(N)$, south $(S)$, and east $(E)$ sides of the device, respectively. Here $H = 100 \,\mu m$.}
    \label{fig:1}
\end{figure}
The problem considered in this study is schematically shown in Fig.~\ref{fig:1}. It is a microfluidic T junction device that has two converging inlets and one outlet. The lengths of both top $(L_{I,T})$ and bottom $(L_{I,B})$ converging inlets are kept at $6.25H$, whereas the outlet length $(L_{O})$ is fixed at $21H$, where $H \, = 100\, \mu m$. The heights of the device in the inlet and outlet sections are $H$ and $2H$, respectively. Binary non-Newtonian electrolyte fluids with high and low electrical conductivities  enter into the microfluidic T junction through the south $(S)$ and north $(N)$ inlet sections of the device, as schematically shown in Fig.~\ref{fig:1}.   

To investigate the EKI phenomenon, we need to solve the convective-diffusive equation for the electrical conductivity, the Ohmic current conservation equation, and the Navier–Stokes equations. A detailed discussion of all these governing equations has already been presented in the prior studies carried out for simple Newtonian fluids~\cite{lin2004instability,chen2005convective,posner2006convective}. For the sake of completeness, we have written those equations briefly in their dimensional forms as follows:\newline  
Continuity equation:
\begin{equation}
    \frac{\partial u_{j}^{*}}{\partial x_{j}^{*}} = 0
\end{equation}
Momentum equation:
\begin{equation}
    \rho \left( \frac{\partial u_{i}^{*}}{\partial t^{*}} + u_{j}^{*} \frac{\partial u_{i}^{*}}{\partial x_{j}^{*}} \right) = -\frac{\partial p^{*}}{\partial x_{i}^{*}} + \frac{\partial \tau_{ij}^{*}}{\partial x_{j}^{*}} - \rho_{e}^{*}E_{i}^{*}
\end{equation}
Ohmic model equation:
\begin{equation}
 \frac{\partial \sigma^{*}}{\partial t^{*}} + u_{j}^{*}\frac{\partial \sigma^{*}}{\partial x_{j}^{*}} = D_{\text{eff}}\frac{\partial^{2} \sigma^{*}}{\partial x_{j}^{*2}}   
\end{equation}
\begin{equation}
    \frac{\partial (\sigma^{*} E_{j}^{*})}{\partial x_{j}^{*}} = 0
\end{equation}
\begin{equation}
     \frac{\partial (\epsilon E_{j}^{*})}{\partial x_{j}^{*}} = \rho_{e}^{*}
\end{equation}

In the above equations, $\rho$ is the fluid density, $\sigma^{*}$ is the electrical conductivity, $\epsilon$ is the electrical permitivity, $E^{*}$ is the electric field, $\rho_{e}^{*}$ is the charge density, $x^{*}$ is the position, $u^{*}$ is the velocity vector, $t^{*}$ is the time, $p^{*}$ is the pressure, and $\tau^{*}$ is the extra stress tensor. $D_{\text{eff}}$ is the effective diffusivity which can be calculated for a binary and monovalent fully dissociated electrolyte as $\frac{2D_{+}D_{-}}{D_{+} + D_{-}}$. Here $D_{+}$ and $D_{-}$ are the diffusivities of positive and negative ions, respectively. These can be related to the ionic mobility $\mu_{\pm}$ through the Einstein relation as $D_{\pm} = RT m_{\pm}$, where $R$ is the universal gas constant and $T$ is the absolute temperature. The electrical conductivity $(\sigma^{*})$ and charge density $(\rho_{e}^{*})$ are related to the ionic species concentration and valency by the equations $F^{2}\left( m_{+}C_{+}z_{+}^{2} + m_{-}C_{-}z_{-}^{2}\right)$ and $F\left(C_{+}z_{+} - C_{-}z_{-}\right)$, respectively, where $F$ is the Faraday constant, $C_{\pm}$ and $z_{\pm}$ are ionic species concentration and valences, respectively. Note that here $(\,)^{*}$ denotes a dimensional variable.  

For a non-Newtonian power-law fluid, the relation between the extra stress tensor $\tau_{ij}^{*}$ and the rate of deformation tensor $\epsilon_{ij}^{*}$ is
\begin{equation}
    \tau_{ij}^{*} = 2 \eta \epsilon_{ij}^{*}
    \label{power}
\end{equation}
where $\epsilon_{ij}^{*}$ is related to the velocity field as follows
\begin{equation}
    \epsilon_{ij}^{*} = \frac{1}{2} \left( \frac{\partial u_{i}^{*}} {\partial x_{j}^{*}} + \frac{\partial u_{j}^{*}} {\partial x_{i}^{*}}   \right)
\end{equation}
The viscosity $\eta$ in Eq.~\ref{power} is given by the following relation 
\begin{equation}
    \eta = \left( \frac{I_{2}^{*}}{2} \right)^{\frac{( n - 1 )}{2}}
\end{equation}
In the above equation, $n$ is the power-law index, and $I_{2}^{*}$ is the second invariant of the rate of the strain tensor. For a shear-thinning fluid, $n < 1$, whereas $n > 1$ corresponds to a shear-thickening fluid. For a Newtonian fluid, $n = 1$. 
The following scaling variables are used to non-dimensionalize the aforementioned governing equations: position with $H$, velocity with $U_{ev}$, time with $\frac{H}{U_{ev}}$, pressure and extra stress tensor with $m \left(\frac{U_{ev}}{H}\right)^{n}$, electric body force with $ \frac{m}{H} \left(\frac{U_{ev}}{H}\right)^{n}$, electrical conductivity with characteristic conductivity $\sigma_{0}$, electric field with $E_{a}$. Here $m$ is the power-law consistency index, $U_{ev}$ is the electroviscous velocity defined as $\frac{\epsilon E_{a}^{2} H}{m}$ and $E_{a}$ is the apparent applied electric field calculated as $\frac{V_{N} - V_{E}}{L_{O} + L_{I,T}}$, where and $V_{E}$ and $V_{N}$ are the voltages applied at the east and north sides of the device. After performing the non-dimensionalization with these scaling variables, one can obtain the following non-dimensional governing equations:
\begin{equation}
    \frac{\partial u_{j}}{\partial x_{j}} = 0
\end{equation}
Momentum equation:
\begin{equation}
    Re \left( \frac{\partial u_{i}}{\partial t} + u_{j} \frac{\partial u_{i}}{\partial x_{j}} \right) = -\frac{\partial p}{\partial x_{i}}  + \frac{\partial \tau_{ij}}{\partial x_{j}} - \rho_{e}E_{i}
\end{equation}
Ohmic model equation:
\begin{equation}
 \frac{\partial \sigma}{\partial t} + u_{j}\frac{\partial \sigma}{\partial x_{j}} = \frac{1}{Ra_{e}}\frac{\partial^{2} \sigma}{\partial x_{j}^{2}}   
\end{equation}
\begin{equation}
    \frac{\partial (\sigma E_{j})}{\partial x_{j}} = 0
\end{equation}
\begin{equation}
     \frac{\partial (\epsilon E_{j})}{\partial x_{j}} = \rho_{e}
\end{equation}
It can be seen that the present EKI flow phenomenon will be governed by two non-dimensional numbers, namely, the Reynolds number defined for power-law fluids as $Re = \frac{\rho U_{ev}^{2-n} H^{n}}{m}$ and the electric Rayleigh number defined as $Ra_{e} = \frac{U_{ev} H}{D_{\text{eff}}}$. Apart from these two dimensionless numbers, additionally, we have the following three dimensionless numbers, namely, the conductivity ratio defined as $\gamma = \frac{\sigma_{H}}{\sigma_{L}}$, the voltage ratio defined as $V_{R} = \frac{V_{S}}{V_{N}}$, and the power-law index $n$.    

\section{\label{sec:Num}Numerical details}
\subsection{Computational solution procedure}
All the governing equations, namely, mass, momentum, Ohmic, and Oldroyd-B viscoelastic constitutive equations, written in the preceding section have been solved using the finite-volume method (FVM) based rheoEFoam solver available in the recently developed RheoTool package~\cite{rheoTool}. This solver has been developed based on the open-source computational fluid dynamics (CFD) code OpenFOAM~\cite{wellerOpenFOAM}. A detailed description of the present solver used in this study is already available elsewhere~\cite{pimenta2018numerical}, and hence only some of the salient features (mainly different discretization techniques) of this solver are recapitulated here. All the advective terms in the governing equations were discretized using the high-resolution CUBISTA (Convergent and Universally Bounded Interpolation Scheme for Treatment of Advection) scheme for improved iterative convergence properties. All the diffusion terms in the governing equations were discretized using the second-order accurate Gauss linear orthogonal interpolation scheme. All the gradient terms were discretized using the Gauss linear interpolation scheme. The backward time integration scheme was used to discretize the time derivative terms. While the linear systems of the pressure, velocity and electric potential fields were solved using the Geometric-Algebraic Multi-Grid (GAMG) with DIC (Diagonal-based Incomplete Cholesky) preconditioner, the stress, dye concentration, and electrical conductivity fields were solved using the Preconditioned Bi-conjugate Gradient Solver (PBiCG) solver with DILU (Diagonal-based Incomplete LU) preconditioner. The pressure-velocity coupling was accomplished using the SIMPLE method, and the log-conformation tensor approach was used to stabilize the numerical solution. Furthermore, the relative tolerance level for the pressure, velocity, stress, dye concentration, and electrical conductivity fields was set as 10$^{-10}$. The whole computational domain was discretized using a total of 57000 hexahedral cells. This number was fixed after performing the standard grid independence study at the lowest value of the power-law index considered in this study.  

Finally, the following boundary conditions have been employed in order to facilitate the numerical simulations. For the pressure, a zero $(p = 0)$ condition at all inlet and outlet sides of the device as they are open to atmosphere and a zero gradient $(\frac{\partial p}{\partial n_{i}} = 0)$ at all solid walls are imposed, where $n_{i}$ is the unit outward normal vector. For the electric potential, a constant value $(\psi = C)$ at all inlet and outlet sides of the device depending upon the values of the apparent electric strength $E_{a}$ and the voltage ratio $V_{R}$ and a zero gradient $(\frac{\partial \psi}{\partial n_{i}} = 0)$ at all solid walls are applied . For the velocity, a zero gradient $(\frac{\partial u_{i}}{\partial n_{i}} = 0)$ at all inlet and outlet sides and a slip boundary condition of the form $u_{s,i} = \mu_{0}\frac{\sigma}{\sigma_{0}}^{m}E_{i}$ are implemented~\cite{chen2005convective}. Here $\mu_{0} = -\frac{\epsilon \zeta_{0}}{\eta_{0}}$ (where $\zeta$ is the wall zeta potential) is a reference electroosmotic mobility at a reference electrical conductivity $\sigma_{0}$ and $m$ is an exponent used to account the power-law dependence of the wall zeta potential on the electrical conductivity. A value of $m = -0.3$ is used in this study as suggested by the literature~\cite{chen2005convective}.

\subsection{Dynamic mode decomposition (DMD) technique}
%%%%%%%%%%%%%%%%%%%%%%%%%%%%%%%%%%%%%%%%%%%%%5
In the present work, we utilize data-driven dynamic mode decomposition algorithm \cite{schmid2010dynamic, schmid2011applications, schmid2011application} to dissect dynamically important subdomains of the flow field into 'modes' associated with characteristic timescales.  To begin with, for the results to converge and capture relevant dynamics (both fast and slow), a sufficient number of snapshots need to be sampled at a high frequency. In this context, we vectorized 1600 ($M$) snapshots ($\bm{s}$) of the concentration field ($N$ degrees of freedom), obtained at a temporal spacing of 0.005s, in the form given by
      \[  {S}_{1}^{M}=   \begin{bmatrix} \bm{s_1}\,\,\, \bm{s_2}    \cdots\,\,\,\bm{s_M} \end{bmatrix} \in \mathbb{R}^{N\times M}\]
here the subscript and superscript of ${S}$ represent the initial and final time steps, respectively. Further, DMD assumes a linear mapping approximation between the two subsequent snapshots of this matrix, as shown below
 \begin{equation} \label{eq1_dmd}
 \bm{s_{j+1}}=\boldsymbol{A}\bm{s_{j}}
 \end{equation}
Moreover, $\boldsymbol{A} \in \mathbb{R} ^{N\times N}$ is approximately constant over the whole sequence, and, therefore, two temporally shifted sets of snapshots are constructed from the parent matrix ${S}_{1}^{M}$, which are related as
  \begin{equation}
   \label{eq2_dmd}
 S_{2}^{M}=\boldsymbol{A}\,S_{1}^{M-1}
 \end{equation}
\par Based on Koopman theory \cite{rowley2009spectral}, the underlying non-linear dynamics is captured by the eigenvectors (Ritz vectors) and eigenvalues (Ritz values) of this system matrix $\boldsymbol{A}$. The Ritz vectors, also referred to as the DMD modes, are numerical approximations of Koopman modes and represent the spatial features of the analyzed flow field, whereas the Ritz values provide information about the associated temporal dynamics.  However, in practice, a companion matrix $ C \in \mathbb{R}^{(M-1)\times (M-1)}$ is built which is a low dimensional representation of $\boldsymbol{A}$,
  \begin{equation}
  \label{eq19}
 S_{2}^{M} \approx \boldsymbol{C}\,S_{1}^{M-1}\,
 \end{equation}
Here, $\boldsymbol{C}$ reflects some of the eigenvalues and eigenvectors of $\boldsymbol{A}$, which are obtained by employing the robust singular value decomposition (SVD) method as follows
\begin{equation}
 \label{eq_dmd_3}
 S_{1}^{M-1}= U\Sigma W^{T}   
\end{equation}
where $U$ is the left unitary matrix, $\Sigma$ is a diagonal matrix comprising singular values, and superscript $^{T}$ denotes the transpose.
 Due to the low rank of the singular matrix, only leading singular values are retained to minimize the noise. Substituting Eq.~\ref{eq_dmd_3} in Eq.~\ref{eq19} and subsequent rank truncation yields $\overset{\boldsymbol{\mathtt{\sim}}}{C}$ as
 \begin{equation}
  \overset{\boldsymbol{\mathtt{\sim}}}{C}= U^{T} S_{2}^{M}W\Sigma^{-1}
\end{equation}
 This is followed by the eigendecomposition of $\overset{\boldsymbol{\mathtt{\sim}}}{C}$ matrix, given by
\begin{equation}
\overset{\boldsymbol{\mathtt{\sim}}}{C} \boldsymbol{q_{j}}= \lambda_{j} \boldsymbol{q_{j}} 
\end{equation}
where $\boldsymbol{q_{j}} \in  \mathbb{R} ^{r} $ are the eigenvectors and corresponding eigenvalues ($\lambda_{j}$) are utilized to obtain $\sigma_{j}$=Re$(log(\lambda_{j})/ \Delta t)
$ and $\omega_{j}$=Im$(log(\lambda_{j})/ \Delta t)$. Here, $\sigma_{j}$ is the growth rate, and $\omega_{j}$ is the angular frequency of $j^{th}$ mode. The dynamic modes $\phi_{j}$ are, therefore, obtained by 
\begin{equation}
\phi_{j}= U\boldsymbol{q_{j}}
\end{equation}
The dynamics of the flow field can be reconstructed as
%%%%%%%%%%%%%%%%%%%%%%%%%%%%%%%%%%%%%%%%%%%%%%%%%%%%%%%%%%%%%%%%%%%%%%%%%%%%%%%%%%%%%%%%%%%%
\begin{align*}
  S_{1}^{M-1} &= \underbrace{[\phi_{1}\,\phi_{2}\,\cdots \,\phi_{r}]}_{\boldsymbol{\Phi\,(Modes)}}
  \underbrace{\begin{bmatrix}
    b_{1}& \\
    &b_{2}& \\
    && \ddots & \\
    &&&& b_{r}
  \end{bmatrix}}_{\boldsymbol{D_{b}\,(Amplitudes)}}
 \underbrace{ \begin{bmatrix} 
    1 & \lambda_{1} & \dots & \lambda_{1}^{M-2} \\
    1 & \lambda_{2} & \dots & \lambda_{2}^{M-2} \\
    \vdots & \vdots & \ddots & \vdots \\
    1 &  \lambda_{r} &  \dots     & \lambda_{r}^{M-2}
    \end{bmatrix}}_{\boldsymbol{V_{and}\,(Dynamics)}}
    \end{align*} 
%%%%%%%%%%%%%%%%%%%%%%%%%%%%%%%%%%%%%%%%%%%%%

where the matrix $V_{and}$
is known as the Vandermonde matrix formed using the eigenvalues of $\overset{\boldsymbol{\mathtt{\sim}}}{C}$ and contains the information regarding temporal dynamics of the modes, $D_{b}$= diag(\textbf{b}) comprises the amplitude of each mode, obtained from $\bold{b}=\Phi^{\dagger}\bold{x_{1}}$, where $^{\dagger}$ represents the Moore-Penrose pseudoinverse. When combined together, these give time coefficient matrix, $Co=\boldsymbol{D_{b}}\,\boldsymbol{V_{and}}$. The energy contribution (relative importance) of each mode can be quantified either by its amplitude $b_{j}$ or by its norm $||\phi_{j}||$.

\section{\label{sec:Res}Results and discussion}
The present study performs simulations for a range of values of the power-law index, $0.8 \le n \le 1.2$. The variation in $n$ is selected in such a way that the influence of shear-thinning $(n < 1)$ and shear-thickening $(n > 1)$ fluid properties can be discussed on the instability and subsequent mixing phenomena in comparison to that seen in a Newtonian $(n = 1)$ fluid behaviour. Furthermore, simulations were carried out at fixed values of the electrical conductivity ratio of $\gamma = 10$, applied electric field strength of $E_{a} = 20000$ V/m and voltage ratio of $V_{R} = 1$. The values of the electric Rayleigh $(Ra_{e})$ and Reynolds $(Re)$ numbers are kept constant at 2771.6 and 2.77, respectively. Note that all the results are presented here for the outlet section of the device, wherein all the flow phenomena occur.  

\begin{figure}[h]
    \centering
    \includegraphics[trim=0cm 0cm 0cm 0cm,clip,width=9cm]{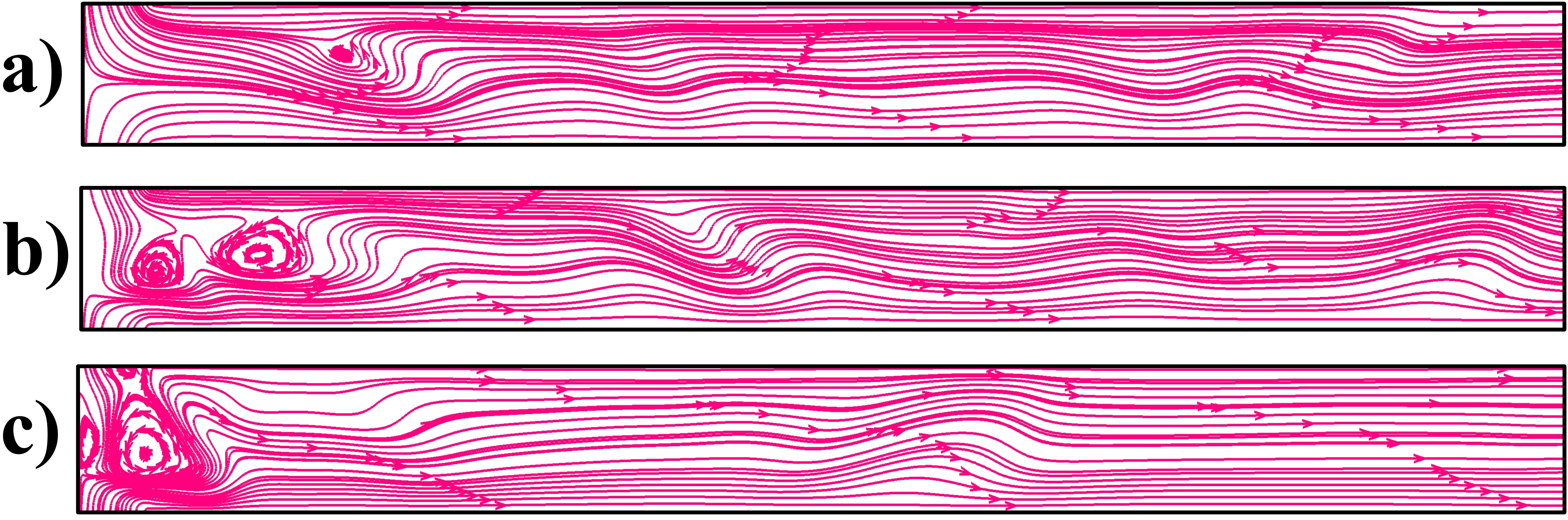}
    \caption{Representative instantaneous streamline patterns inside the microdevice for three values of the power-law index, namely, (a) $n = 1.2$, (b) $n = 1.0$ and (c) $n = 0.8$.}
    \label{fig:stream}
\end{figure}

First, the streamlines inside the microdevice are presented in Fig.~\ref{fig:stream} for three different values of the power-law index, namely, 0.8, 1, and 1.2. This is an efficient post-processing method to visualize the flow pattern inside a system, but limited to a two-dimensional flow field. Here high and low electrical conductivity fluids enter the microdevice from the south and north inlets, respectively. They meet at the junction of two inlets placed at the origin and ultimately leave through the east side of the outlet section of the microdevice. This can also be visualized from the instantaneous streamline patterns presented in Fig~\ref{fig:stream}. The streamlines are seen to be distorted in the outlet section of the device instead of placed in parallel for all values of $n$. This suggests the presence of EKI instability at the interface of the two fluids in the outlet section of the device. A careful inspection reveals that this distortion in the streamline patterns is more in shear-thinning fluids (sub-Fig.~\ref{fig:stream}(c)) as compared to that seen in shear-thickening (sub-Fig.~\ref{fig:stream}(a)) and Newtonian fluids (sub-Fig.~\ref{fig:stream}(b)). In particular, the flow field is seen to be more distorted at the entry of the outlet section where EKI instability starts to originate. This ultimately leads to more chaotic behaviour in the flow field for shear-thinning fluids than for shear-thickening and Newtonian fluids. 

\begin{figure}
    \centering
    \includegraphics[trim=0cm 0cm 0cm 0cm,clip,width=8.5cm]{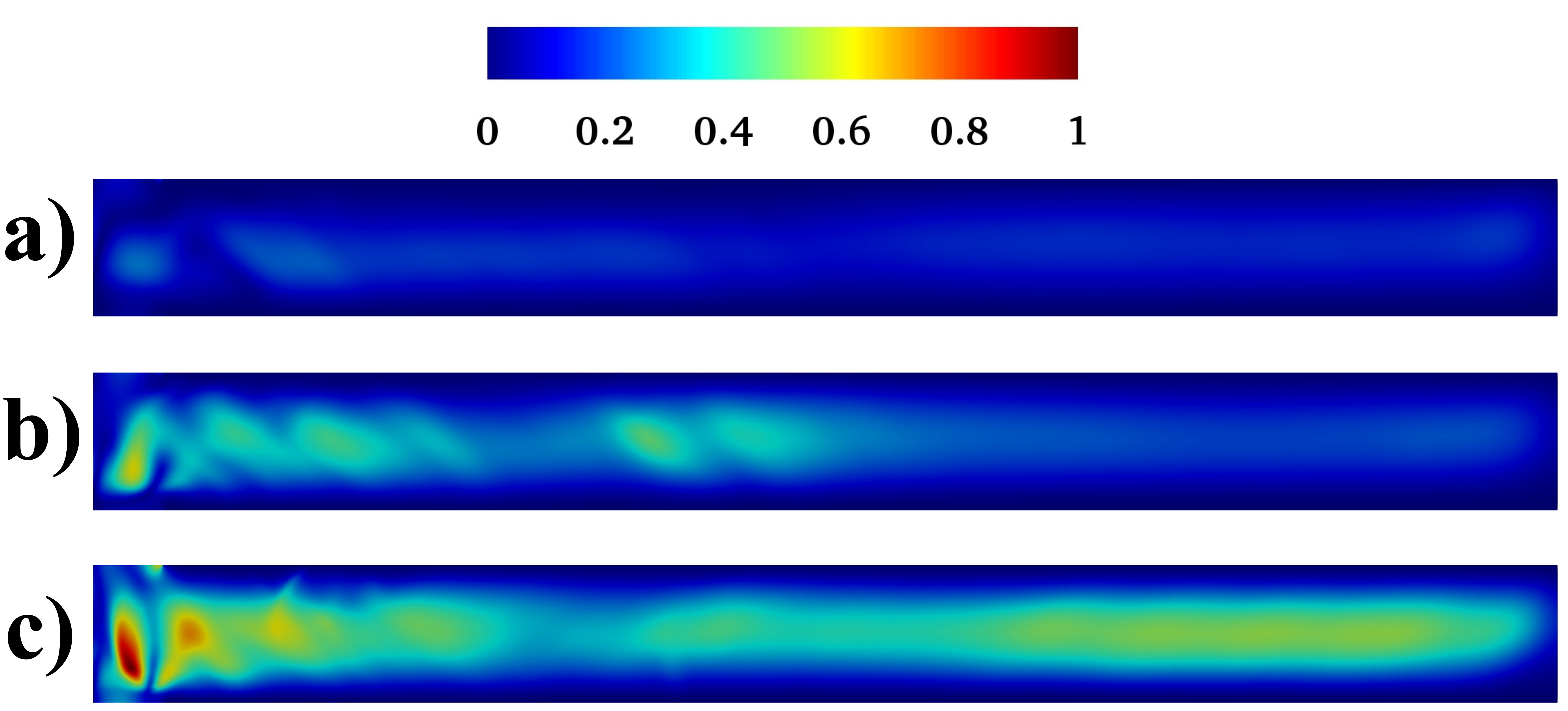}
    \caption{Variation of the time-averaged root mean square span-wise velocity fluctuation $u_{rms,2}$ inside the microdevice for three values of the power-law index, namely, (a) $n = 1.2$, (b) $n = 1.0$ and (c) $n = 0.8$.}
    \label{fig:u2prime}
\end{figure}

\begin{figure}[h]
    \centering
    \includegraphics[trim=0cm 0cm 0cm 0cm,clip,width=8.5cm]{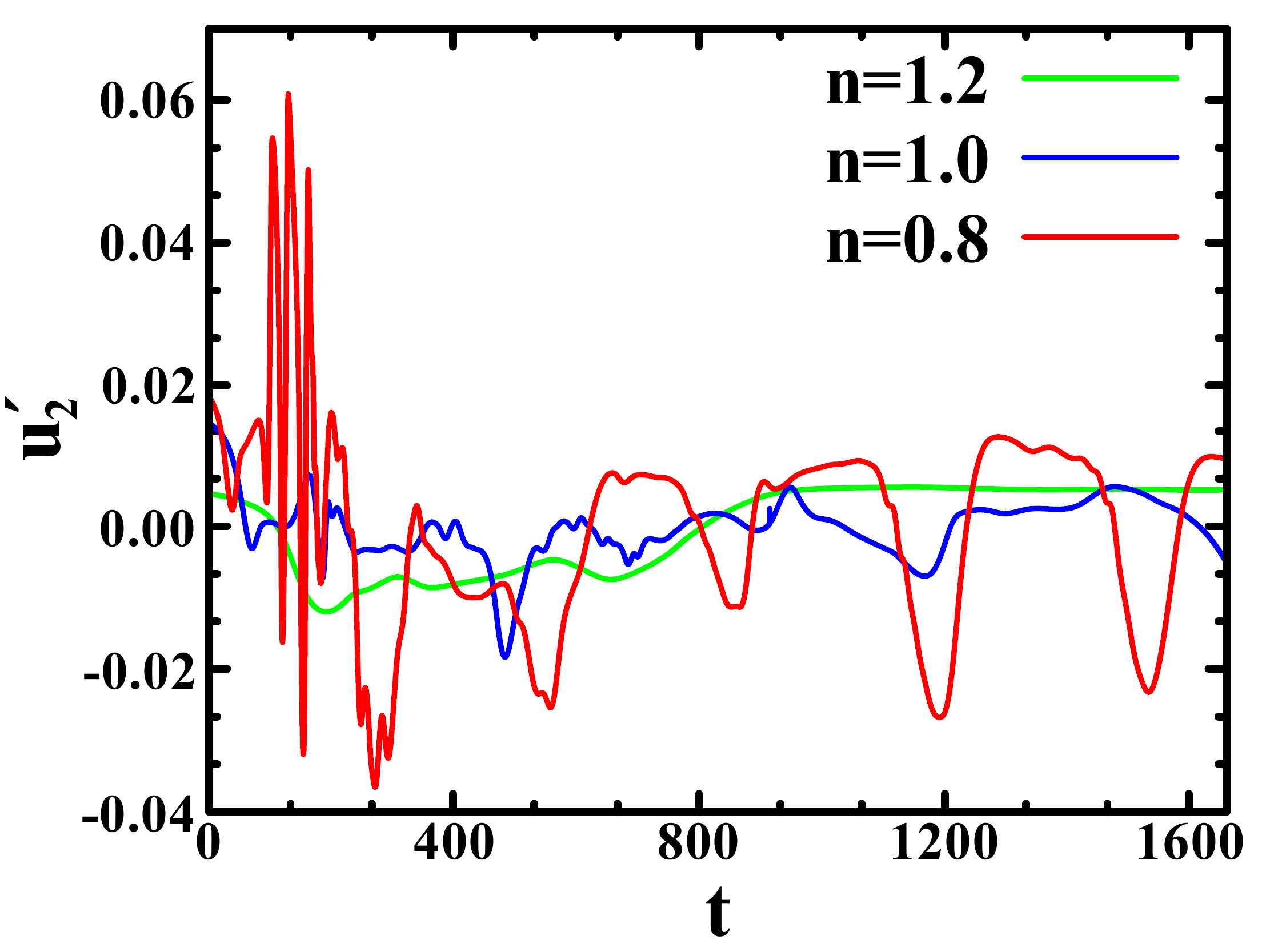}
    \caption{Variation of the span-wise velocity fluctuation $u_{2}^{'}$ inside the microdevice for different values of the power-law index at a probe location $x_{1} = 0.25H $ and $x_{2} = 0$.}
    \label{fig:u2primeProbe}
\end{figure}

This chaotic behaviour originating from the EKI instability in the flow field is ultimately related to the velocity fluctuations. Figure~\ref{fig:u2prime} represents the surface plot of the time-averaged root mean square span-wise velocity fluctuation $u_{rms,2} \,\,(= \sqrt{<(\tilde{u_{2}} - \bar{u}_{2})^{2}>_{t}})$, where $\tilde{u}_{2}$ is the instantaneous span-wise velocity and $\bar{u}_{2}$ is its time-averaged value. It is clearly evident from this figure that the span-wise velocity is seen to be more fluctuating in nature for shear-thinning fluids compared to shear-thickening and Newtonian fluids. This will ultimately result in more mixing of these fluids, which will be discussed in detail later in this section. To present these fluctuations more quantitatively, we have plotted the temporal variation of the span-wise velocity component fluctuation at a probe location ($x_{1} = 0.25H,  x_{2} = 0$) placed at the outlet section of the microdevice, Fig.~\ref{fig:u2primeProbe}. The fluctuation in the span-wise velocity component is found to be more for shear-thinning fluids, whereas it is less for shear-thickening fluids, particularly at the beginning when the instability starts.    

\begin{figure}
    \centering
    \includegraphics[trim=0cm 0cm 0cm 0cm,clip,width=8.5cm]{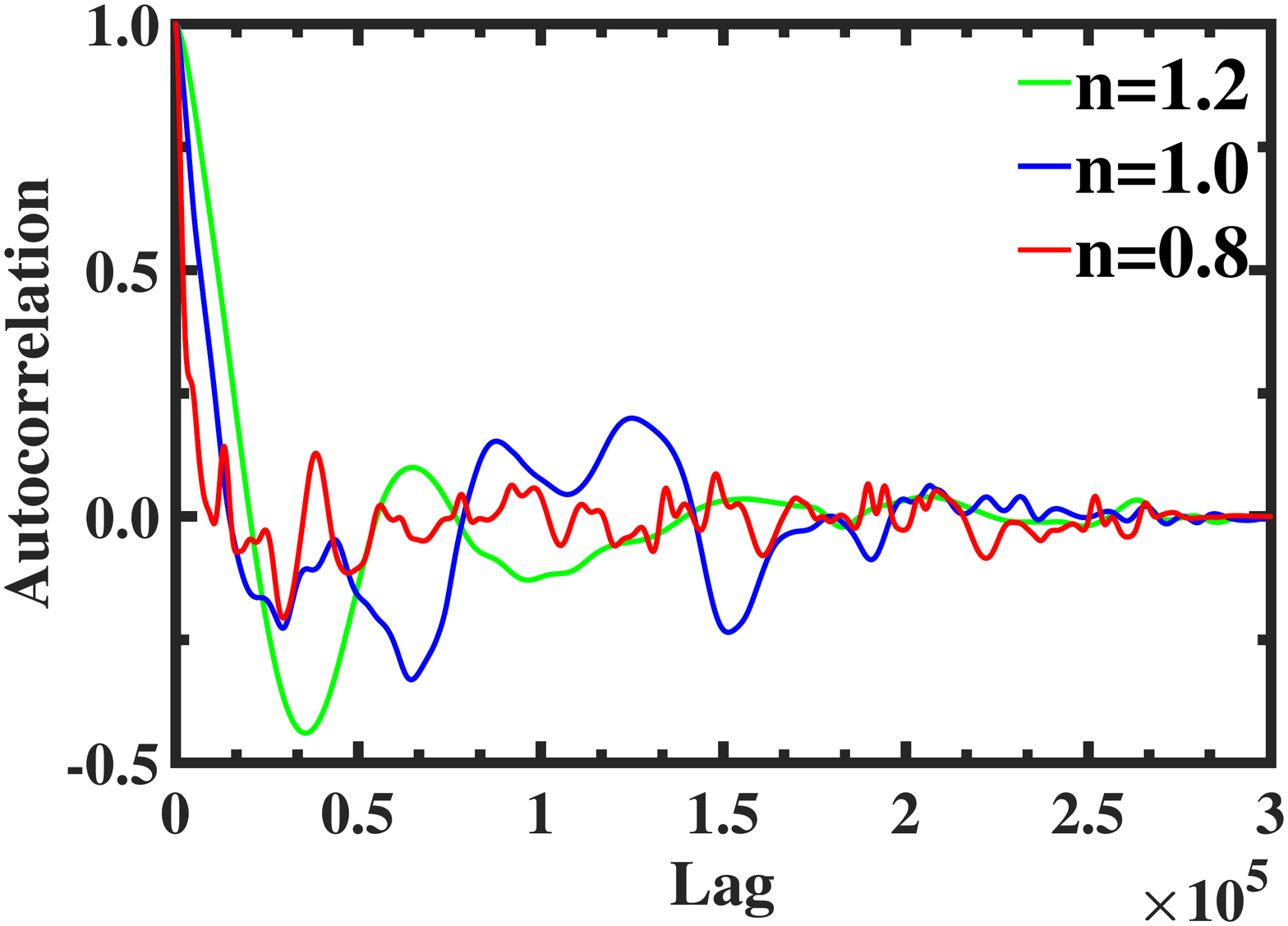}
    \caption{Autocorrelation for the span-wise velocity component variation at a probe location $x_{1} = 5H $ and $x_{2} = 0$ inside the microdevice for different values of the power-law index.}
    \label{fig:autoC}
\end{figure}

\begin{figure}
    \centering
    \includegraphics[trim=0cm 0cm 0cm 0cm,clip,width=9cm]{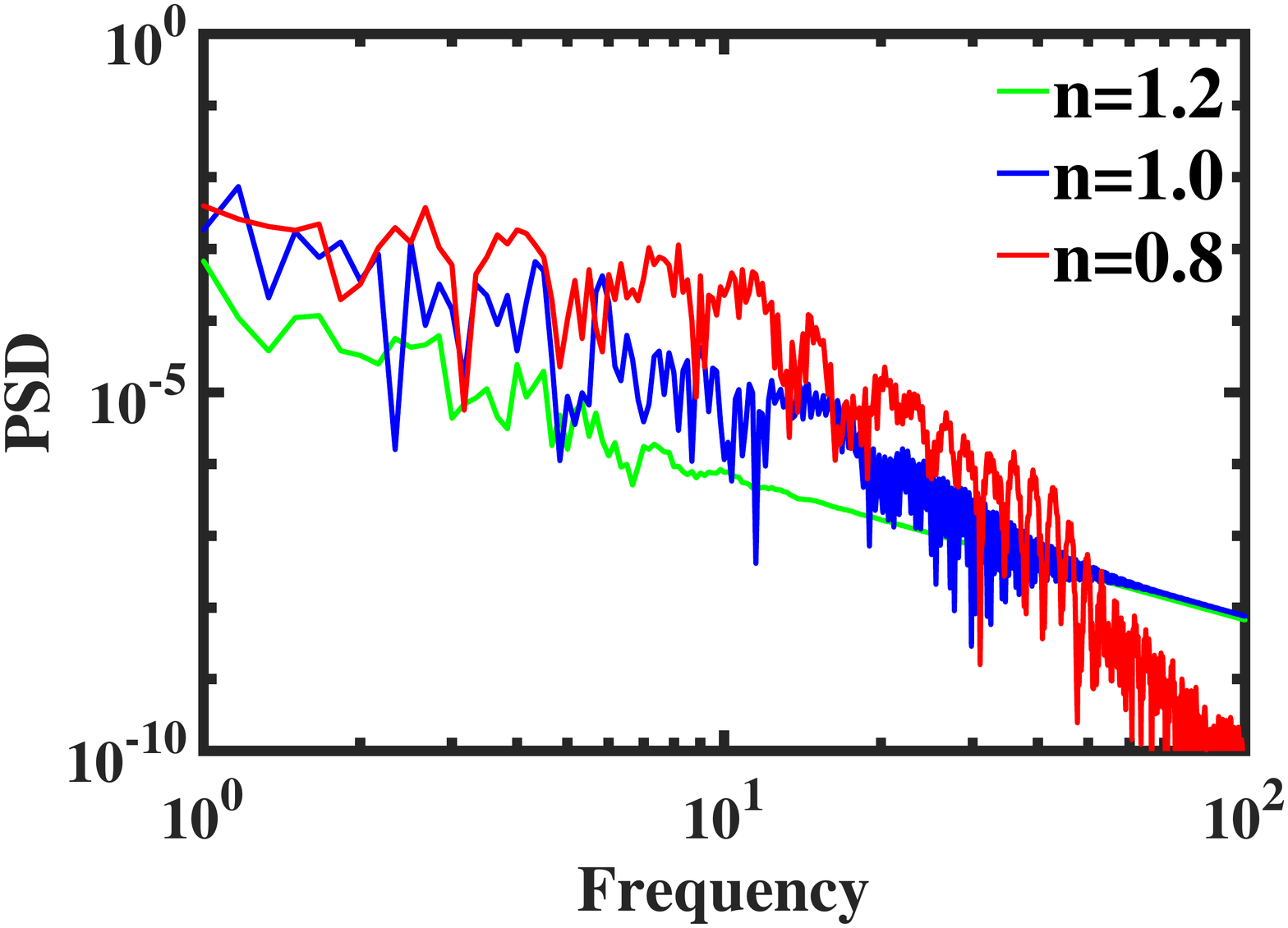}
    \caption{Power spectral density plot of span-wise velocity component fluctuation at a probe location $x_{1} = 0.25H $ and $x_{2} = 0$ inside the microdevice.}
    \label{fig:PSD}
\end{figure}

To obtain more statistical insights into the chaotic flow dynamics inside the microdevice, we have plotted the autocorrelation of the span-wise velocity component variation at three different values of the power-law index in Fig.~\ref{fig:autoC}. When the lag time is small, a steep decrease in the correlation value is observed as the value of $n$ gradually decreases, indicating that the variation of the span-wise velocity component becomes gradually less correlated as we move from shear-thickening to shear-thinning fluids via the Newtonian one. This suggests that the randomness in the velocity fluctuation is more in shear-thinning fluids compared to that seen in shear-thickening and Newtonian fluids under otherwise identical conditions. Over a large value of the lag time, there is no correlation present for all kinds of fluids. On the other hand, Figure~\ref{fig:PSD} depicts the corresponding power spectrum of the span-wise velocity component fluctuations for different values of the power-law index. It can be readily seen from this figure that for shear-thinning fluids, the velocity field is excited over a broad range of time and length scales in comparison to that seen either for shear-thickening or Newtonian fluids. Also, the PSD magnitude is higher for shear-thinning fluids, indicating a larger intensity of velocity fluctuations for these fluids. 
%%%%%%%%%%%%%%%%%%%%%%%%%%%%%%%%%%%%%%%%%%%%%%

All the aforementioned results presented herein suggest that the intensity of electrokinetic instability gradually increases as the fluid rheological behaviour gradually changes from shear-thickening to shear-thinning via the Newtonian one. This behaviour can be, at least, qualitatively explained as follows: for power-law fluids, the effective viscosity in a flow system scales as $\sim\,u^{n-1}$ where $u$ and $n$ are the velocity and power-law index, respectively. As a result, the effective viscosity gradually tends to decrease as we decrease the value of the power-law index or we move towards the shear-thinning behaviour of the fluid. Therefore, the effective viscosity is expected to be lower in shear-thinning fluids than in shear-thickening fluids under the same conditions. This decrease in the effective viscosity in shear-thinning fluids $(n < 1)$ leads to more span-wise (as well as stream-wise) fluctuation in the velocity field due to the lowering in the viscous forces, which have a tendency to suppress the instability. This, in turn, accelerates the EKI phenomenon more in shear-thinning fluids. On the other hand, the effective viscous forces increase in shear-thickening fluids $(n > 1)$ and suppress the velocity fluctuations in these fluids.   

\begin{figure}[h]
    \centering
    \includegraphics[trim=0cm 0cm 0cm 0cm,clip,width=9cm]{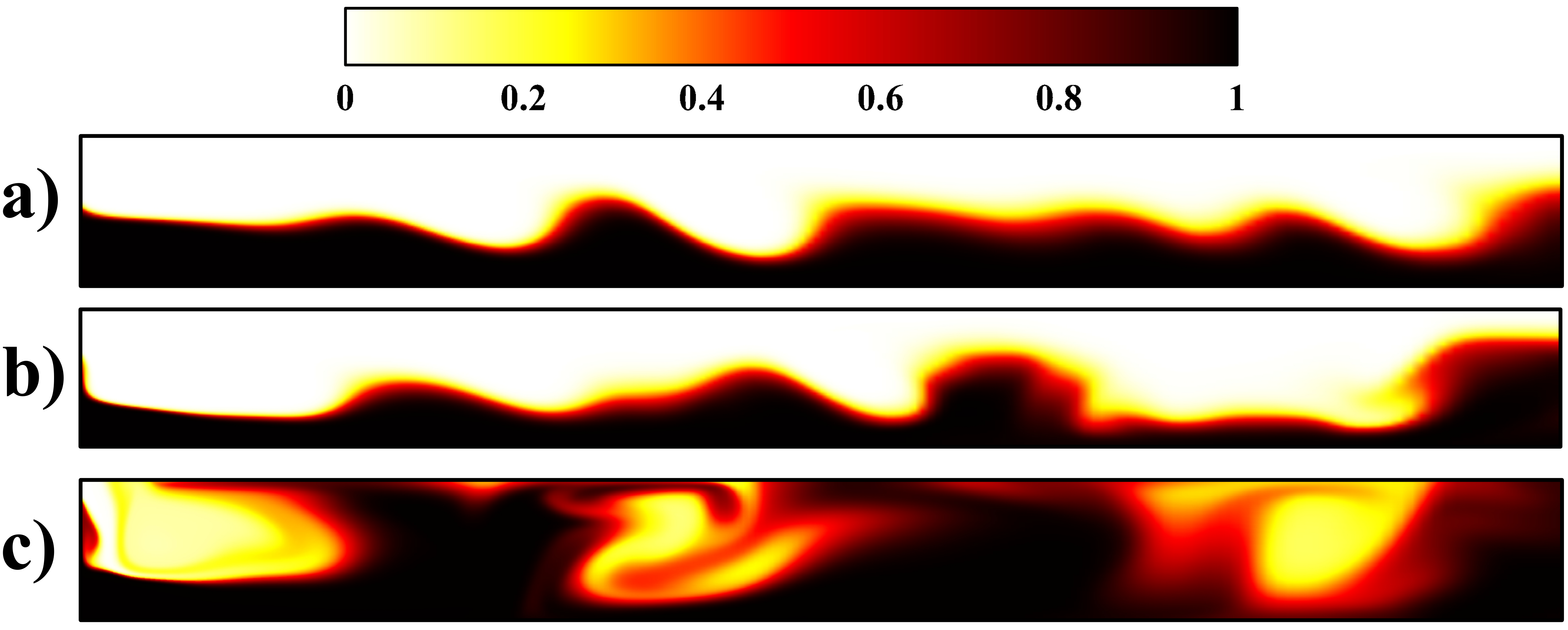}
    \caption{Instantaneous dye concentration profile inside the microdevice for different values of the power-law index. (a) $n = 1.2$, (b) $n = 1$, and (c) $n = 0.8$.}
    \label{fig:C_Ins}
\end{figure}

It is now expected that the mixing process, often performed utilizing this EKI phenomenon, would also be greatly influenced by the fluid rheological behaviour in the present microdevice. To analyze this, we first present the instantaneous dye concentration profile inside the microdevice in Fig.~\ref{fig:C_Ins}. 

\begin{figure}
    \centering
    \includegraphics[trim=0cm 0cm 0cm 0cm,clip,width=9cm]{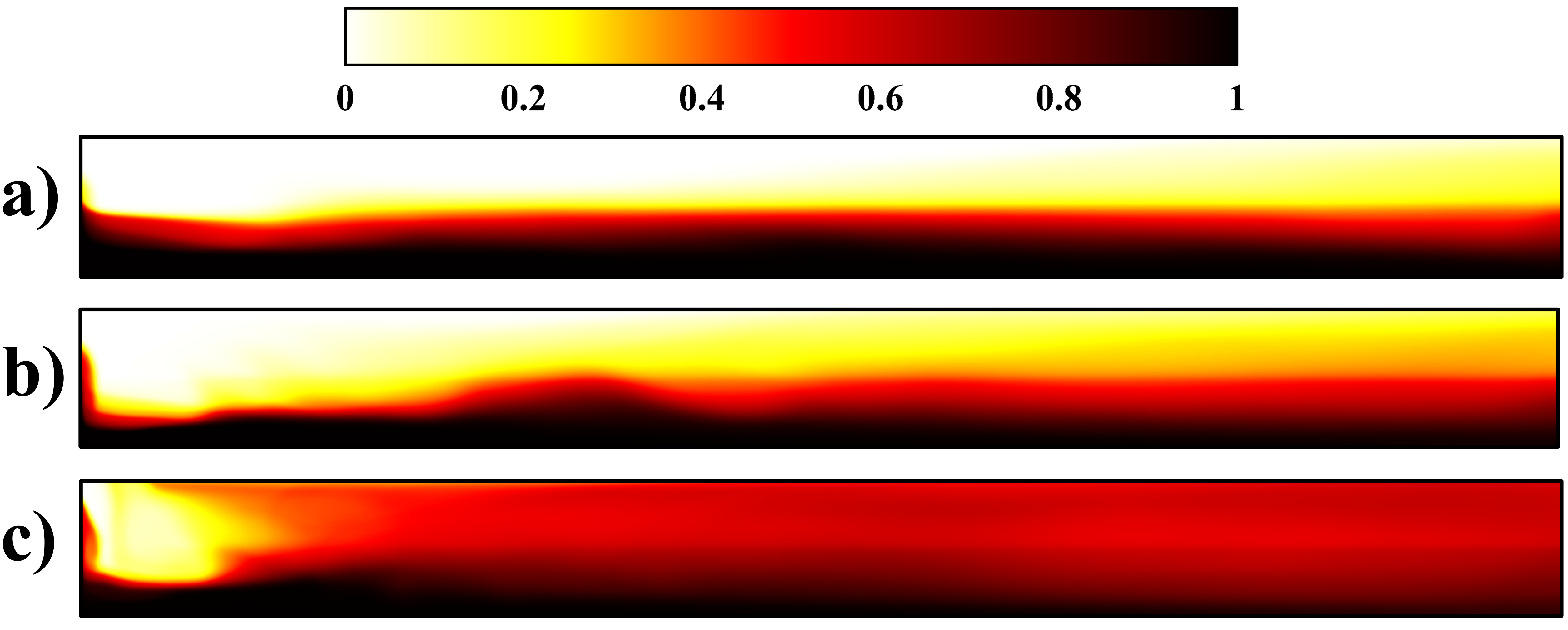}
    \caption{Time-averaged mean dye concentration profile inside the microdevice for different values of the power-law index. (a) $n = 1.2$, (b) $n = 1$, and (c) $n = 0.8$.}
    \label{fig:C_Mean}
\end{figure}

\begin{figure}
    \centering
    \includegraphics[trim=0cm 0cm 0cm 0cm,clip,width=9cm]{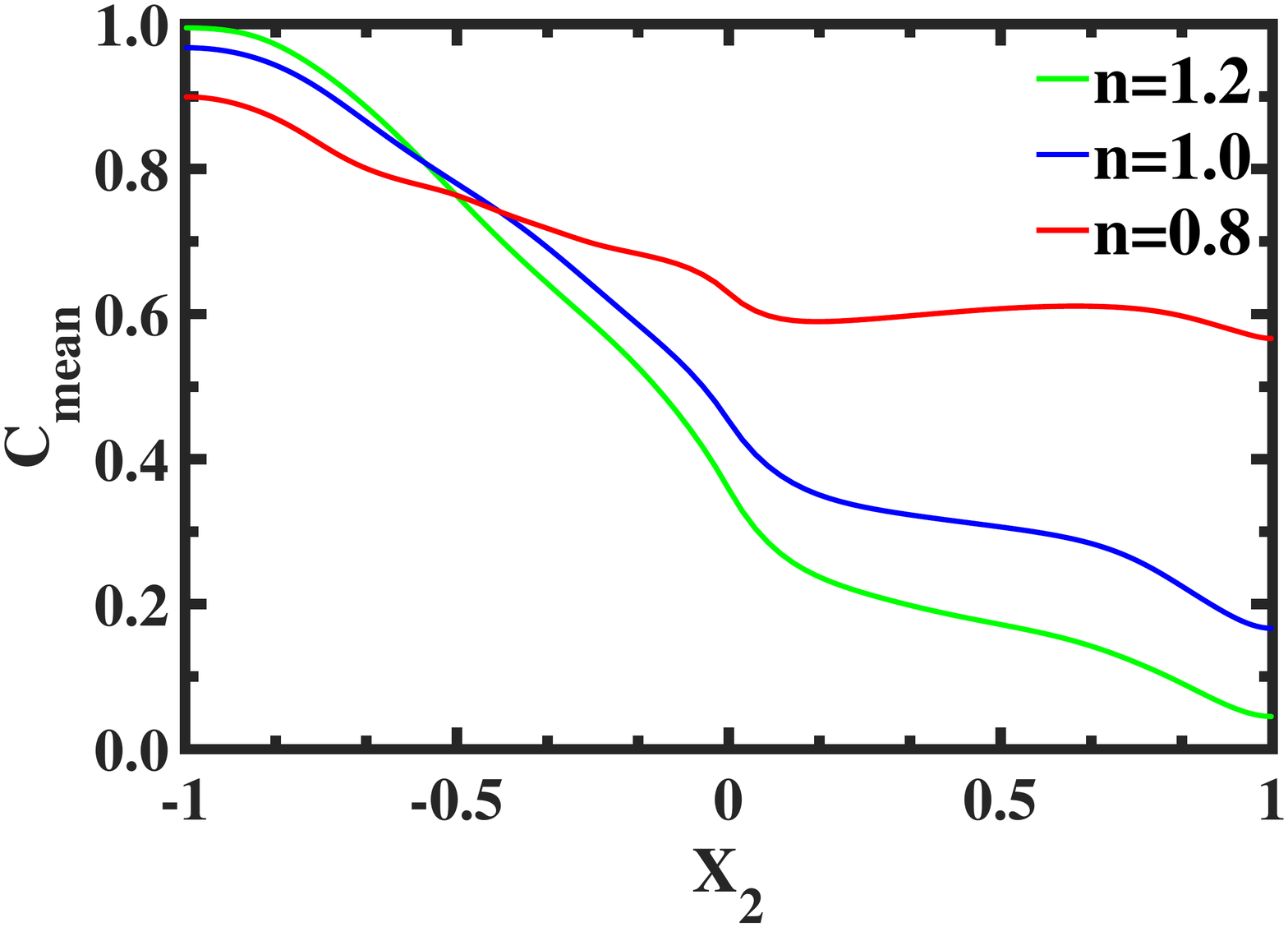}
    \caption{Variation of the time-averaged mean dye concentration along the span-wise direction at the outlet place of the microdevice for different values of the power-law index.}
    \label{fig:C_MeanVsPosi}
\end{figure}

Note that here fluids entering into the microdevice from the south and north inlets have finite non-dimensional dye concentrations of $c = 1$ and 0, respectively. The following convective-diffusive equation of the form $\frac{\partial c}{\partial t} + u_{j} \frac{\partial c}{\partial u_{j}} = \frac{1}{Pe}\frac{\partial^{2}c}{\partial x_{j}^{2}}$ (where $Pe \,( = \, U_{ev}\, H / D)$ is the Peclet number and $D$ is the diffusivity of ions. A value of around 26778 for this number is used in the present study) has been solved for the evaluation of the dye concentration inside the microdevice along with other governing equations. The interface of the two fluids, where the gradient of the dye concentration is maximum, is wavy in nature for shear-thickening (sub-Fig.~\ref{fig:C_Ins}(a)) and Newtonian (sub-Fig.~\ref{fig:C_Ins}(b)) fluids. On the other hand, for shear-thinning fluids (sub-Fig.~\ref{fig:C_Ins}), the dyed fluid travels from the bottom half of the channel to the top half (and vice-versa for undyed fluid), and as a result, the fluid interface can not be tracked. This is because of high chaotic convection inside the device for these fluids, as already discussed above. Therefore, one can expect a greater extent of mixing of two fluids if they are shear-thinning in nature. This is, indeed, observed in Fig.~\ref{fig:C_Mean}, wherein the time-averaged dye concentration $(C_{mean})$ is plotted for fluids with different values of the power-law index. It is clearly noticeable from this figure that the dye is more uniformly distributed in the outlet section of the device for shear-thinning fluids as compared to that seen in shear-thickening and Newtonian fluids. To show it more quantitatively, we have plotted the distribution of $C_{mean}$ along the non-dimensional span-wise direction at the outlet plane of the microdevice in Fig.~\ref{fig:C_MeanVsPosi}. The slope of the variation of $C_{mean}$ curve with the span-wise direction gradually decreases as the fluid behaviour progressively changes from shear-thickening to shear-thinning fluids via the Newtonian one. This suggests that the distribution of dye is more uniform in shear-thinning fluids, resulting from a larger mixing of two fluids present in the top and bottom halves of the microdevice. 


\begin{figure}
    \centering
    \includegraphics[trim=0cm 0cm 0cm 0cm,clip,width=9cm]{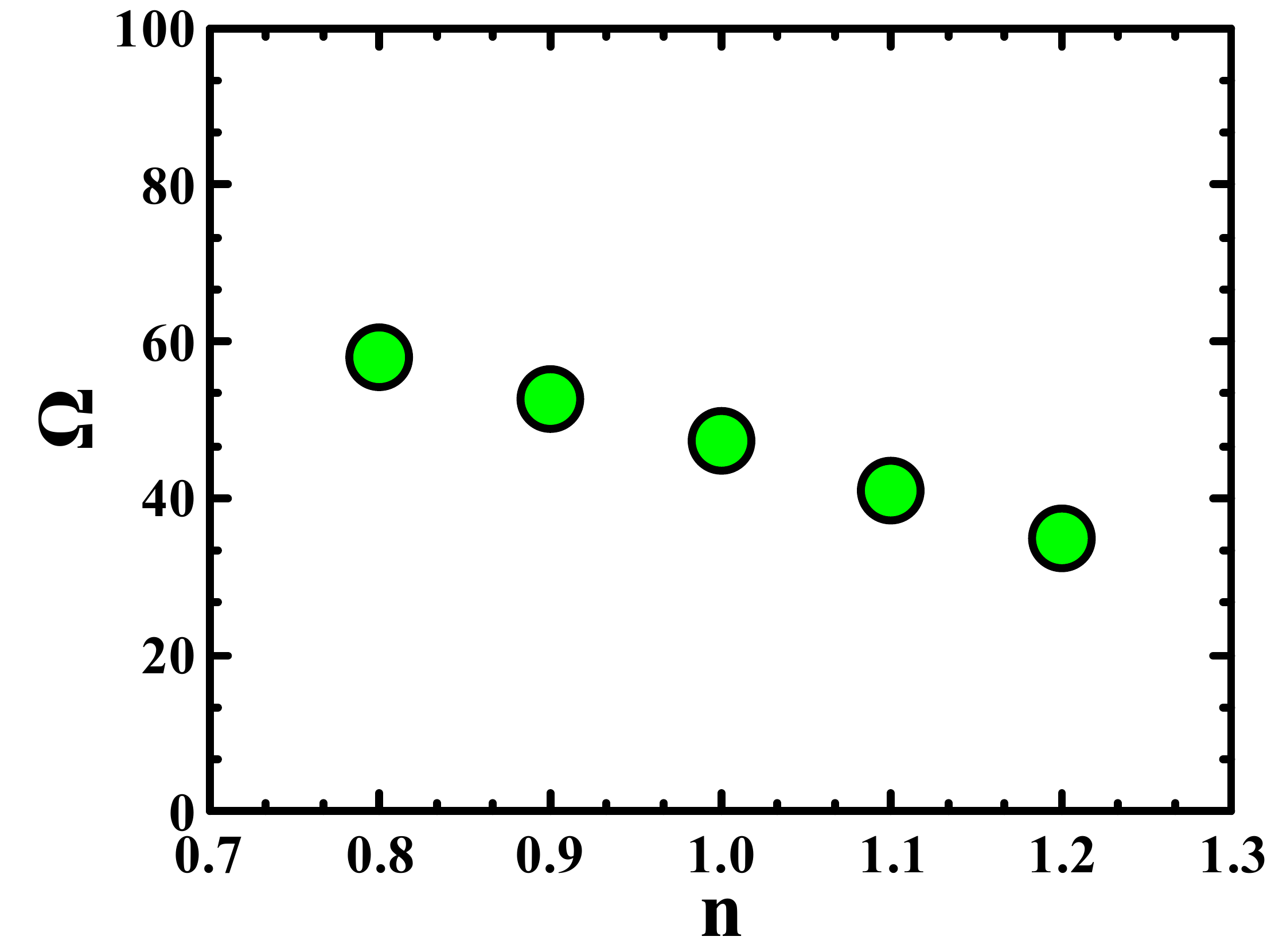}
    \caption{Variation of the mixing efficiency with different values of the power-law index.}
    \label{fig:mixeffi}
\end{figure}

The corresponding mixing efficiency $\Omega$ is calculated to present this mixing phenomenon in a more quantitative manner, which is defined as
\begin{equation}
   \Omega = 1 - \frac{\sqrt{\frac{1}{N}\sum_{1}^{N}(C_{mean}^{N} - C_{mean}^{*})^{2}}}{\sqrt{\frac{1}{N}\sum_{1}^{N}(C_{mean}^{0} - C_{mean}^{*})^{2}}}
\end{equation}
where $C_{mean}^{N}$, $C_{mean}^{0}$ and $C_{mean}^{*}$ are the dye concentration at a point along $x_{2}$ direction at the outlet plane, dye concentration of unmixed fluids and dye concentration of perfectly mixed fluids, respectively. The value of $C_{mean}^{0}$ is either 0 or 1, resulting the value of $C_{mean}^{*}$ as 0.5. The variation of this parameter with the power-law index is illustrated in Fig.~\ref{fig:mixeffi}. It can be seen that $\Omega$ gradually decreases as the value of the power-law index increases, i.e., as the fluid rheological behaviour gradually transits from shear-thinning to shear-thickening behaviour via the Newtonian one due to a lowering in the intensity of the EKI phenomenon.            

\begin{figure*}
    \centering
    \includegraphics[trim=0cm 0cm 0cm 0cm,clip,width=18cm]{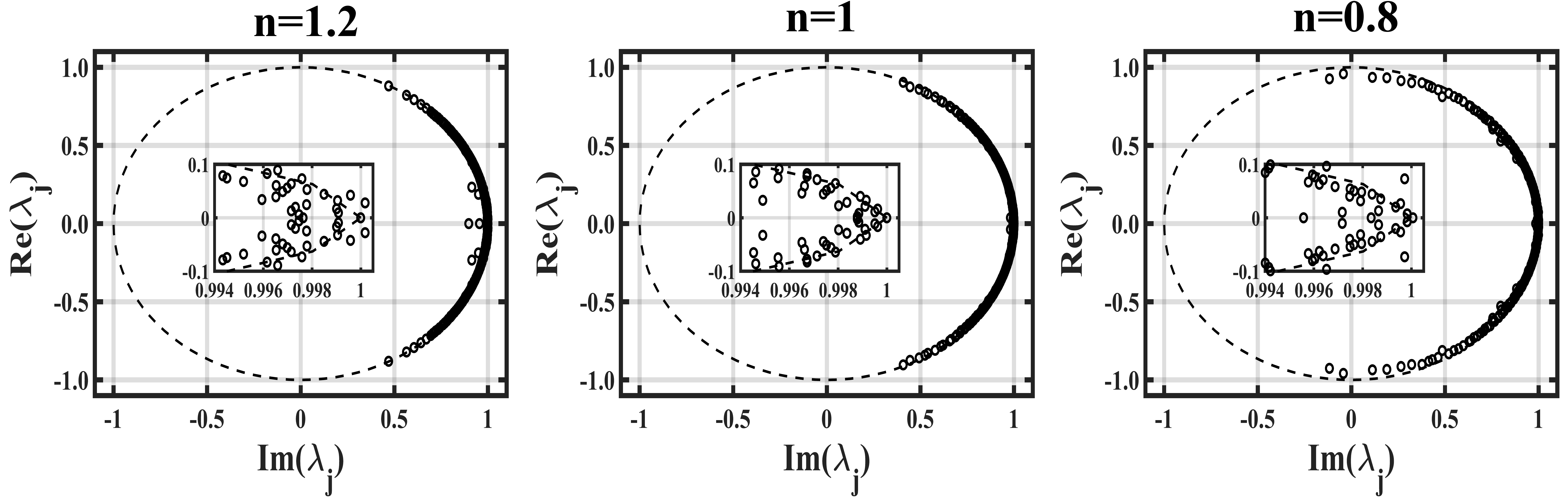}
    \caption{Ritz values $\lambda_{j}$ for shear thinning, Newtonian, and shear thickening fluids, respectively. A close-up of these values near the circumference is presented in the zoomed-in figures.}
    \label{fig:lambda}
\end{figure*}

\begin{figure*}
    \centering
    \includegraphics[trim=0cm 0cm 0cm 0cm,clip,width=18cm]{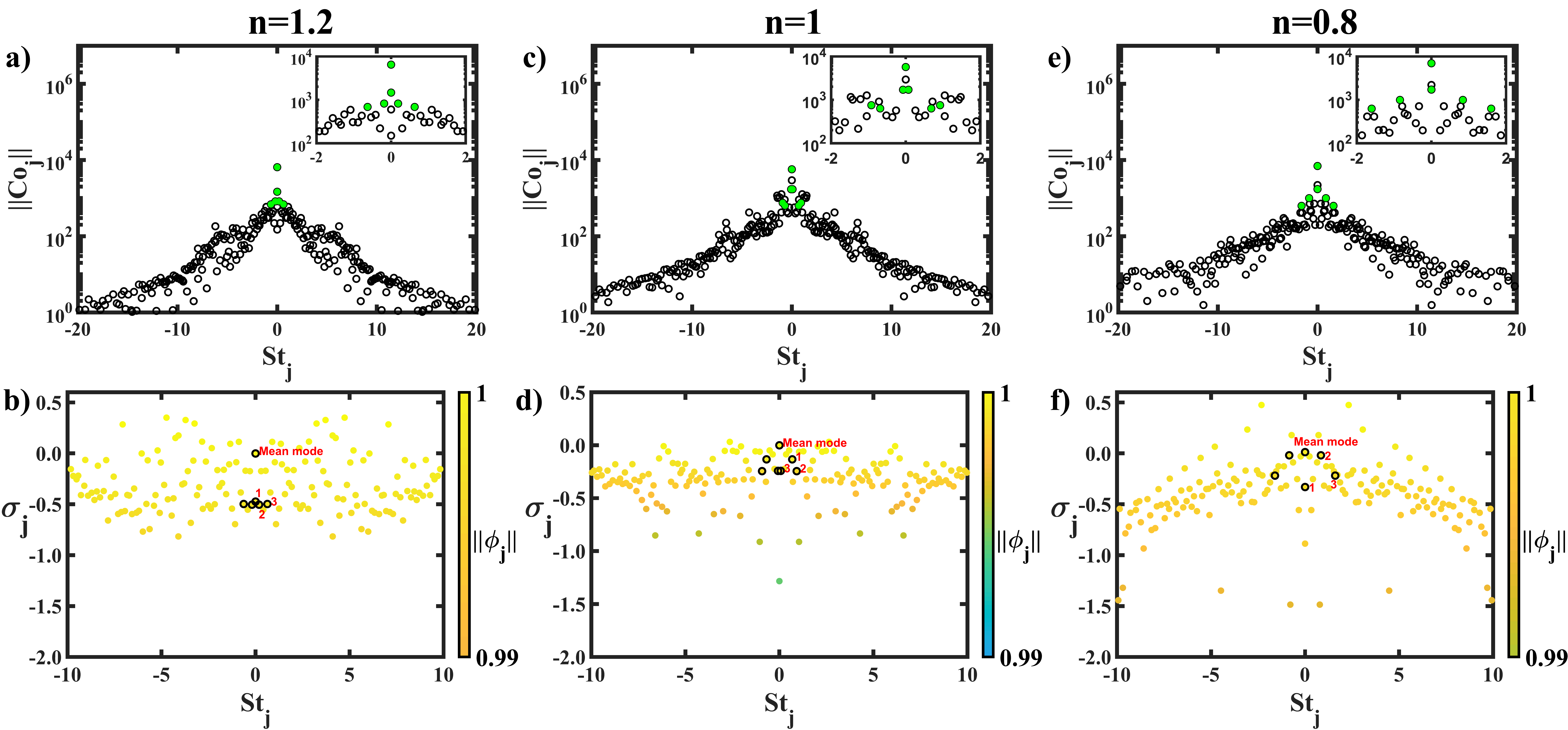}
    \caption{Time coefficient norm ($||Co_j||$) and growth rate ($\sigma_{j}$) against the frequency ($St_{j}$) for: $n = 1.2$ (a, b); $n = 1$ (c, d); and $n = 0.8$ (e, f). The color bar in $\sigma_{j}$ vs $St_{j}$ plot represents the mode norm ($||\phi_j||$). Further, the modes having highest values of $||Co_j||$ are encircled and numbered in order in $\sigma_{j}$ vs $St_{j}$ plot. The unlabeled modes are the corresponding complex conjugates of the chosen modes.}
    \label{fig:combinedandlandscape}
\end{figure*}

\begin{figure*}
    \centering
    \includegraphics[trim=0cm 0cm 0cm 0cm,clip,width=18cm]{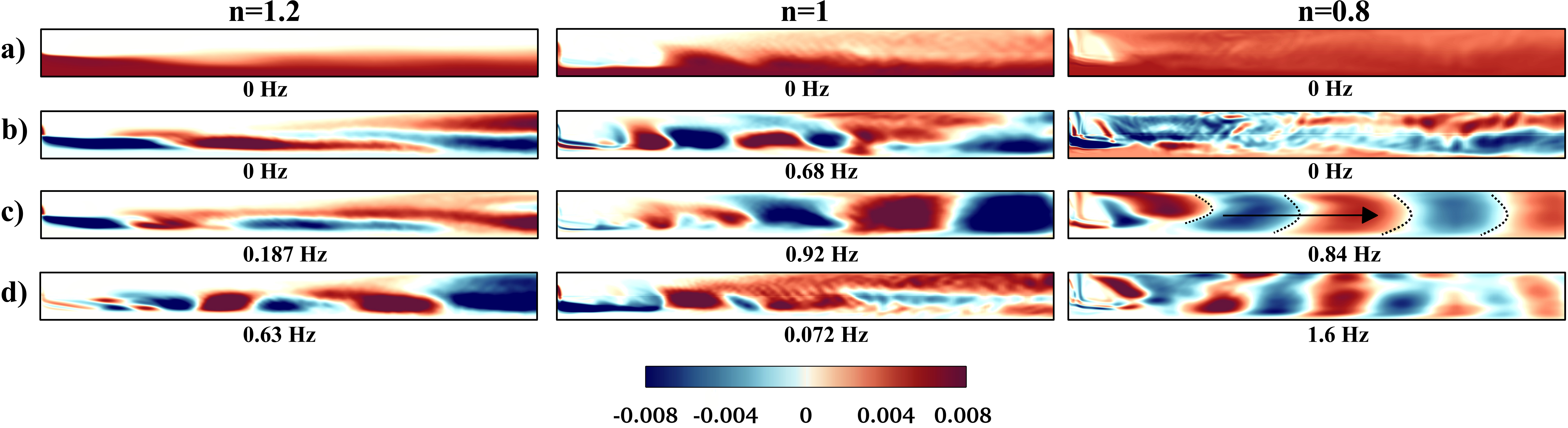}
    \caption{Representative DMD modes inside the microdevice for three values of the power-law index, namely, $n = 1.2$, $n = 1.0$ and $n = 0.8$. Here, (a) corresponds to mean modes, whereas (b), (c), and (d) represent mode 1, mode 2, and mode 3, along with associated frequencies, respectively, in each fluid type.}
    \label{fig:modes}
\end{figure*}

Now, we analyze the underlying coherent structures at different values of the power-law index obtained using the procedure described in section~\ref{sec:Num}. The aim of utilizing this powerful data-driven analysis tool is to better understand the differences that exist in the characteristic flow features of rheologically different fluids as the EKI phenomenon sets in. First, the Ritz values corresponding to the first 300 modes for all three fluids are illustrated in Fig.~\ref{fig:lambda}. 
%%%%%%%%%%%%%%%%%%%%
%%%%%%%%%%%%%%%%%%%%%%%%%%%%%%%%%%%%%%%%%%%%%%%%%%%%%%%%%%%%%%%%%
It can be seen that most of the Ritz values fall either on the unit circle or inside it, while very few lie outside, indicating the presence of quasi-steady state dynamics in the system. This suggests that the EKI phenomenon and the subsequent mixing process are strongly complex and non-linear. Furthermore, many unstable frequencies prevail in the system, which was also evident in the PSD plot presented in Fig.~\ref{fig:PSD}. As a result, only a few neutrally stable structures appear. The DMD modes are often sorted based on the mode amplitude or norm, as already stated earlier. However, in the present study, multiple modes exist with a very slight difference in their energy due to the strongly chaotic nature of the flow system. Therefore, in addition to the mode norm, we resort to another parameter, called the time-coefficient norm $||Co_j||$, for choosing the modes which reflect the dynamics more appropriately. A similar sorting criterion has also been employed by a number of earlier studies \cite{wan2015dynamic, huang2022dynamic}, wherein the driving frequency of the system has been utilized to segregate relevant modes in addition to $||Co_j||$. First, the plots of the growth rate and time coefficient with respect to the frequency are shown in Fig.~\ref{fig:combinedandlandscape} for power-law indices 1.2, 1, and 0.8, respectively. 
%%%%%%%%%%%%%%%%%%%%%%%%%%%%%%%%%%%%%%%%%%%%%%
%%%%%%%%%%%%%%%%%%%%%%%%%%%%%%%%%%%%%%%%%%%%%%
It is clear from the frequency $(St_{j}=\frac{\omega_{j}}{2 \pi})$ versus the growth rate plot that a large number of modes have very slight differences in their energy, as the values of the mode norms vary only between 0.99 and 1, irrespective of the fluid type. Furthermore, for each fluid, there is a mode with zero growth rate and zero frequency corresponding to the mean mode having the highest value of the time-coefficient norm. The mean modes for all three fluids are shown in Fig.~\ref{fig:modes} (a) and are similar to the corresponding mean concentration fields shown in Fig.~\ref{fig:C_Mean}. As concluded from the mean concentration field results, mean DMD modes also depict substantially higher mixing in the shear-thinning fluids than in the Newtonian and shear-thickening fluids. The rest of the modes (having the highest $||Co_j||$ value) are first sorted on the basis of least damping values and then numbered in descending order of $||Co_j||$; see encircled modes in Fig.~\ref{fig:combinedandlandscape}. It is worth mentioning that most of the obtained modes exhibit similar structures at various time scales. Therefore, we have chosen the top three modes in each case to explain the underlying dynamics in the flow system.
%%%%%%%%%%%%%%%%%%%%%%%%%%%%%%%%%%%%%%%%%
%%%%%%%%%%%%%%%%%%%%%%%%%%%%%%%%%%%%%%%%%%%%%%
\par At first glance, irrespective of the fluid type, the structures embedded in these modes exhibit similar cellular patterns, albeit to a varying degree of spatial extent; see Fig.~\ref{fig:modes} (b-d). The existence of such underlying structures can be explained on the basis that the present EKI phenomenon results due to a coupling between electrohydrodynamic instability (EHD) and electroosmotic flow (EOF)\cite{chen2005convective}. Initially, due to the EHD instability at the diffusive interface, slanted perturbations are originated in the presence of electric body forces. This results in the curving and folding of the interface, which in turn drives the high-conductivity fluid from the bottom and the low-conductivity fluid from the top. On the other hand, under the influence of the applied electric field, an accumulation of free charge occurs at these slanted interfaces, causing the formation of separate zones (structures) of positive and negative perturbations in the form of cells. This cellular motion of the fluids further enhances the instability \cite{hoburg1976internal}. Thereafter, the EOF plays a convective role, which advances the instability to the right side of the micro-channel and ensures the large-scale mixing of the fluids. After taking a closer look at these modes, it can be seen that the structures are distributed along the interface. This makes it clear that the instability initiates at the interface and not in the electric double layer, as already emphasized by earlier studies also \cite{chen2005convective}. In the case of shear-thickening fluid, first, two modes are nearly alike, with alternate positive (red) and negative (blue) structures placed at the interface between the two fluids and stretched in the horizontal direction. The frequencies associated with the first (b) and second (c) modes are 0 and 0.187 Hz, respectively, indicating that these basic and large structures change slowly over time. On the other hand, the structures in the third mode (d) are smaller in size and are not stretched that much, and they also evolve at a higher frequency of 0.63 Hz. Furthermore, the expanse of these structures in the span-wise direction increases as the fluid traverses from the entrance to the exit of the outlet duct. This further supports the explanation that the instability originates at the interface near the entrance of the outlet section, and thereafter, the EOF convects the amplifying disturbance with some frequency towards the end of the outlet section. On comparing the structures of rheologically different fluids, a significant difference can be seen. The first substantial difference is seen to be present in the expanse of these structures in the span-wise direction. The shear-thinning fluids exhibit the highest expanse from the entrance of the outlet section itself. Comparing the lowest frequency (large) structures in all three fluids, i.e., 0 Hz (mode 1 (b)) for $n$ = 1.2 and 0.8, and 0.072 Hz (mode 3 (d)) for ${n}$ = 1, it is quite clear that the shear-thinning DMD structures cover nearly the whole outlet section of the microdevice. It is followed by the Newtonian fluid, for which the width of these structures is restricted to the interfacial region, while for the shear-thickening fluid, this entrance length is extended further downstream. This observation supports our earlier statement that the intensity of the EKI increases as the fluid behaviour gradually passes from shear-thickening to shear-thinning via the Newtonian one. These wider DMD structures ultimately result in more mixing in the shear-thinning fluids than in the shear-thickening and Newtonian fluids. Moreover, the intensity of these structures for shear-thickening (e.g., mode 3 (d)) and Newtonian fluids (e.g., mode 3 (d)) remains the same throughout the channel, whereas, for shear-thinning fluids (e.g., mode 2 (c)), the intensity abates gradually. A possible reason for this trend is that the mixing phenomenon for the former two fluids slowly starts from the entrance of the outlet section and intensifies towards the end. In contrast, for the shear-thinning fluids, severe mixing takes place at the entry of the outlet section itself, and as a result, the conductivity gradient reduces downstream of the outlet section, which is also reflected from the DMD structures. Furthermore, the convection of the underlying coherent structures can also be easily visualized in mode 2 (c) (0.84 Hz) of shear-thinning fluid, where under the stress of a tangential electric field, the resulting EOF generates a rightward motion in these structures and causes some distortion in them. The structures in mode 3 (d) are smaller and move at nearly double the frequency as that of mode 2 (c), which highlights the typical feature of the DMD, wherein it extracts the higher harmonics which depict the convection of finer structures.

\section{\label{sec:con}Conclusions}
This study presents a numerical investigation of the influence of the fluid rheological behaviour on the electrokinetic instability (EKI) and subsequent mixing phenomena in a microfluidic T-junction. The non-Newtonian power-law model with different values of the power-law index $(n)$ has been used to realize the rheological behaviour of the fluid. It has been found that the intensity and chaotic convection arising due to this EKI phenomenon is higher in shear-thinning fluids $(n < 1)$ than that in Newtonian $(n 
 = 1)$ and shear-thickening fluids $(n > 1)$. As a result, the subsequent mixing phenomenon, which is often achieved using this EKI phenomenon in many micro total analysis systems $(\mu$TAS), has also been found to be significantly enhanced in shear-thinning fluids compared to that in Newtonian and shear-thickening fluids. A possible explanation for this behaviour and a detailed analysis of the results has been presented in this study. Furthermore, the data-driven dynamic mode decomposition (DMD) technique has been utilized to obtain in-depth information on the coherent flow structures at different values of the power-law index. This has ultimately facilitated a better understanding of the differences in the chaotic flow dynamics and mixing process for different power-law fluids and also helped to explain many observations, such as why shear-thinning fluids show higher mixing efficiency than Newtonian and shear-thickening fluids. A significant difference in the expanse and intensity of these coherent flow structures was observed as the fluid rheological behaviour changed.          
\nocite{*}
\bibliography{aipsamp}% Produces the bibliography via BibTeX.

\end{document}